# FendOff encryption software to secure personal information on computers and mobile devices

Victor Solovyev[1] and Ramzan Umarov[2]

[1]*Softberry Inc., Mount Kisco, USA*
[2]*Department of Computer Science, King Abdullah University of Science and Technology Thuwal, Kingdom of Saudi Arabia*
victor.softberry.com, ramzan.umarov@kaust.edu.sa

Abstract: The paper describes several original cryptographic cipher modules (VSEM) that are based on using one time pseudorandom pad and pseudorandom transpositions. The VSEM includes 4 modules of encryption that can be applied in combinations. We studied ability of these modules to secure the private data against attacks and their speed of encryption. The VSEM encryption was implemented in Fendoff applications for mobile devices on iOS and Android platforms as well as in computer application running Window or Mac OS. We describe these applications designed to encrypt/decrypt various personal data such as passwords, credit card or bank information as well as to secure content of any text or image files.

## 1 INTRODUCTION

Computers and networking have entered almost every aspect of human society. They provide hugely increased data transfer over the Internet or other media with large portion of private or personal information in the form of numerous passwords, credit cards and bank account information, customer and medical records content and etc. Security of such data is extremely important and is one of the most active fields of research.

High-quality ciphers and advanced software need to be developed to provide timely protection against the data thieves' attacks. Popular storage of data files in the cloud just increases these security requirements. As billions mobile/smart phones are used on a regular basis and they are currently becoming the universal devices to access various services and data, it is important to develop data protection applications not only for computer systems, but also for mobile communication devices.

The focus of this article is to present relatively fast and robust Various Secure Encryption Modules (VSEM) as well as to analyse their performance. We also describe our several publicly available data security applications for Android and iOS mobile devices as well as for Mac and Window desktop computers.

## 2 VSEM: VARIOUS SECURE ENCRYPTION MODULES

### 2.1 Previous studies

A lot of encryption algorithms are based on usage of pseudo-random sequences (Schneier, 1996). Stream cipher converts stream of data to ciphertext by operating on it bit by bit using a key stream generator (random number generator, for example), and a mixing function, which is often a XOR function. The security provided by these algorithms critically depends on the quality of random sequence generators. Poor random numbers may produce correlations that may be used by cryptanalyst to attack the system. Any computationally generated random sequences are periodic (Schneier, 1996), therefore for practical usage we need to use ones with very long sequence's period to generate a unique finite sequence of random numbers.

## 2.2 Encryption modules

We utilize a combination of symmetric stream and block encryption modules using pseudo-random number generators (R1, R2, ..,Rn), which use (S1, S2,…Sn) seed values to define their initial states setting a generator to the corresponding starting point. Suppose we have a text or image size L bytes (plaintext). The simplest way to transform it to the ciphertext is to apply XOR function to each byte of the plaintext file (**evsem_x** module). This procedure can be combined with any other module described below or embedded to its algorithm.

### 2.2.1 VSEM transposition module

This module selects i-th byte of the plaintext, generate random number j on interval (i + 1, L – 1) and exchange i-th and j-th bytes if j-th byte was not already used in exchanges. If j-th byte was used then i-th byte is exchanged with the first unused byte after byte j and in case of its absence first unused byte before it. The Java code of the encryption module **evsem_t** is following.

```java
byte[] evsem_t (byte[] bm, long seed) {
    int len = bm.length int ip;
    byte bc = 0;
    Rand3 rnd = new Rand3();
    rnd.setSeed(seed);
    boolean[] bb = new boolean[len];
for (int i = 0; i < len; i++)
     bb[i] = true;
for (int i = 0; i < len - 1; i++) {
    if (bb[i]) {
    ip = (int)rnd.rand(i + 1, len - 1);
    bc = bm[ip];
    bm[ip] = bm[i];
    bm[i] = bc;
    bb[ip] = false;
} else if ((ip + 1) < len) {
  ip++;
  while (!bb[ip]          ) {
  ip++;
  if (ip >= len) break; }
  if (ip < len)         {
    bc = bm[ip];
    bm[ip] = bm[i];
    bm[i] = bc;
    bb[ip] = false  }} else     ip--;
    while (!bb[ip] && (ip > i)) {
  if (ip >= len) break; ip--;}
  if (ip <= i) break;}}}
    return bm; }
```

The transpositions itself produce random pattern (see Figure 1), but should be complemented by XOR module to make balance of black and white colors in a case of b/w photo.

The Java code of the decryption module **dvsem_t** is below.

```java
byte[] dvsem_t(byte[] bm, long seed) {
  int len=bm.length;   int ip;
  byte bc = 0;
  Rand2 rnd = new Rand2();
  rnd.setSeed(seed);
  boolean[] bb = new boolean[len];
for(int i=0;i<len;i++)bb[i]=true;
for(int i=0;i<len-1;i++)   {
  if(bb[i]){
  ip= rnd.rand(i+1,len-1);
if(bb[ip]){ bc=bm[ip]; bm[ip]=bm[i];
  bm[i]=bc; bb[ip]=false; }
else if ((ip+1) < len){ ip++;
while (!bb[ip]) {ip++;
 if(ip >=len)break;}
 if(ip < len){ bc=bm[ip]; bm[ip]=bm[i];
   bm[i]=bc; bb[ip]=false;}}
else { ip--; while (!bb[ip] && (ip> i))
{if(ip >=len)break;ip--;}
 if(ip <= i)break;    }}}
 return bm;}
```

### 2.2.2 VSEM shifting module

The shifting **evsem_s** module makes circular shifts of byte content defined by random number for each byte. It has the following encryption code.

```java
byte[] evsem_s (byte[] bm, long seed)
    {
   int len = bm.length;
   byte rb, tb;
   Rand1 rnd = new Rand1();
   rnd.setSeed(seed);
  for (int i = 0; i < len; i++) {
    int j = (int) rnd.rand(0, 256);
    rb = bm[i];
    tb = (byte) (( (rb>>3) ^ (j<<3)));
    bm[i] =  (byte) (rb ^ tb);      }
   return bm; }
```

The corresponding decryption is following.

```java
byte[] dvsem_s(byte[] bm, long seed) {
   int len = bm.length;
byte rb, tb; Rand1 rnd = new Rand1();
   rnd.setSeed(seed);
   for (int i = 0; i < len; i++) {
   byte jj = (byte) rnd.rand(0, 255);
    int j = (int) rnd.rand(0, 7);
    rb = (byte) (bm[i]^jj);
bm[i]  =    (byte)( (0xFF & rb)>>> (8-j)
   | ( rb<<j ) ); }
   return bm; }
```

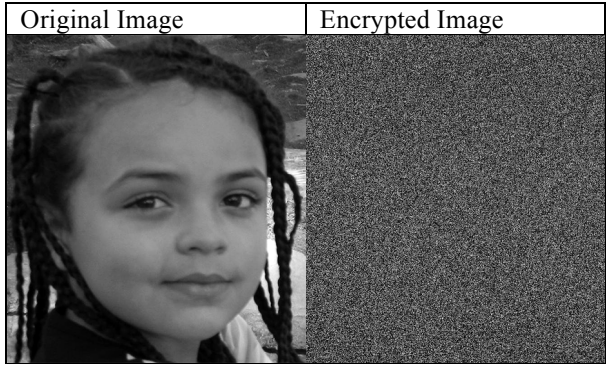

**Figure 1.** Image encrypted by **evsem_t** module.

### 2.2.3 VSEM Circular transposition

The circular transposition module is designed to prevent easy alignment of encrypted texts. Encryption is done by `evsem_ct` procedure.

```
byte[] evsem_ct (byte[] bm, long seed)
    {
  int len=bm.length; byte rb;
  byte bms [] = new byte [len];
  Rand3 rnd = new Rand3();
  rnd.setSeed(seed);
  int jl= (int) rnd.rand(0,len-1);
for(int i=0;i<len;i++){
  if(i+jl < len) bms[i+jl]=bm[i];
  else bms[i+jl-len]=bm[i];}
for(int i=0;i<len;i++)    {
  byte j= (byte) rnd.rand(-128,127);
    rb=(byte) bms[i];
bm[i]=   (byte) (rb^j);}
return bm;  }
```

Decryption is encoded in `dvsem_ct` module.

```
byte[] dvsem_s (byte[] bm, long seed){
    int len = bm.length;
    byte rb, tb;
    Rand1 rnd = new Rand1();
    rnd.setSeed(seed);
    for (int i = 0; i < len; i++) {
    byte jj = (byte) rnd.rand(0, 255);
     int j = (int) rnd.rand(0, 7);
       rb = (byte) (bm[i]^jj);
       bm[i] =    (byte)( (0xFF &
       rb)>>> (8-j) | ( rb<<j ) );;}
  return bm;  }
```

### 2.3 Random generators and their seeds setting

Encryption of a plaintext is secured by the password that is provided by the user and known only to the user. We use the password to define the seed values for our random generators. We divide the password into four parts (p1, p2, p3, p4) and increment initial seed value specific for a particular random generator (R1, R2, …). For example, if we use R1 we have its seed equal to 0xCAFEBCDE+p1. For very short passwords we just increment them by some constant numbers.

Good random generators should do very well on tests of randomness and have periods large enough suitable for most applications. A set of simple and extremely fast RNGs have been developed by combining xorshift operations (Marsaglia, 2003). We used several of them in our encryption modules. The following code presents the typical core part of such generators.

```
public long rand() {
   seed ^= (seed << 13);
   seed ^= (seed >>> 7);
   seed ^= (seed << 17);
return seed; }
```

Using different random generators in our encryption modules provides much longer period of random sequence. Combinations of VSEM modules also increase ciphertext resistance to a cryptanalyst attack. In our applications we apply the following configuration. Encryption: evsem_x -> evsem_t -> evsem_s -> evsem_cs and the corresponding decryption: dvsem_xs -> dvsem_s -> dvsem_t -> dvsem_x.

## 3 PERFORMANCE ANALYSIS

There are several metrics that have been developed (Hamid, 2014) which is often used in testing encryption algorithms (Arul, Venkatesulu, 2012; Reddy, Kumar, 2014). We should note that most of these metrics itself estimate only particular aspects of encryption quality, therefore they might serve only indicators of it, but not as the absolute measure.

### 3.1 Encryption Quality (EQ)

To test encryption quality we used a set of images of various sizes. We apply here the EQ measure and image correlation analysis. The encryption quality measure estimates deviation a grey scale pattern of encrypted image from the original one. The higher the value of EQ, the more the encrypted image is deviated from the original image. Good encryptions will have large deviations. The EQ calculation formula is (Arul, Venkatesulu, 2012):

$$EQ = \frac{\sum_0^{255}|H_L(F') - H_L(F)|}{256}$$

Here $H_L(F)$ is the array of number of occurrences of each grey level $L$ in original image $F$ and $H_L(F')$ is the array of number of occurrences of each grey level in encrypted image.

The test results are presented in Figure 2 where we give original and encrypted images and their EQ values for our three VSEM encryption modules. It can be seen that there is no visual information observed in the encrypted image and they are basically demonstrate the random pattern. The EQ values for our encryption modules are comparable with those computed for other encryption algorithms (Reddy, Kumar, 2014).

| Original Image | Encrypted Image | Block | EQ |
|---|---|---|---|
| 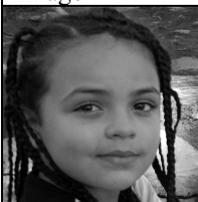 | 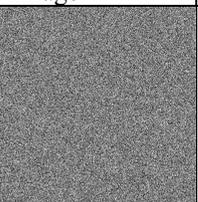 | evsem_t evsem_x | 670 |
| 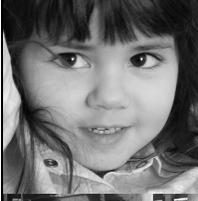 | 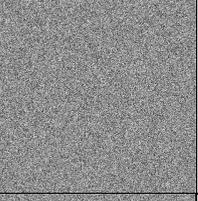 | evsem_s | 583 |
| 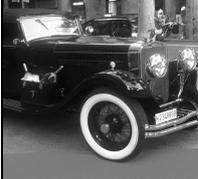 | 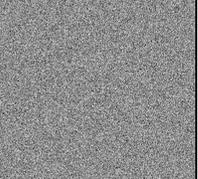 | evsem_ct | 876 |

**Figure 2.** EQ and encrypted images for three VSEM modules

| Original Image | Encrypted Image | EQ | Corr. |
|---|---|---|---|
| 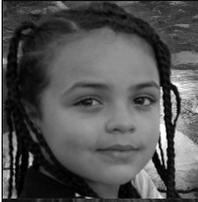 | 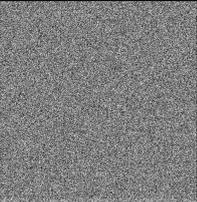 | 669.7 | -0.014 |
| 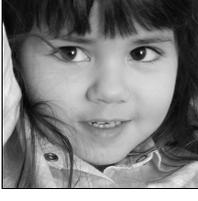 | 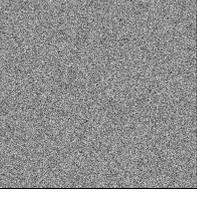 | 582.5 | 0.013 |

**Figure 3.** Encrypted images, EQ and CC after applying all three VSEM modules

### 3.2 Correlation Analysis

If cipherimage is close to random pattern the correlation coefficient computed for the pixel values of two adjacent pixels will be close to zero, and it will be in most cases totally different for the plainimage. The correlation coefficient is computed by the following equation (Krishnamurthy, Ramaswamy, 2009):

$$CC = \frac{Cov(x,y)}{\sqrt{D(x)D(y)}}$$

Where x and y are gray-scale pixel values of the plainimage or cipherimage. Here $Cov(x,y) = \frac{1}{N}\sum_{i=1}^{N}[(x_i - E(x))(y_i - E(y))]$, $D(x) = \frac{1}{N}\sum_{i=1}^{N}[x_i - E(x)]^2$, $D(xy) = \frac{1}{N}\sum_{i=1}^{N}[y_i - E(y)]^2$.

The pair of pixels can be selected in horizontal, vertical, diagonal and anti diagonal directions. For correlation coefficient computation we took 1000 random adjacent pixels for original and encrypted images. The values of CC are presented in Figure 3 and in the adjacency based maps (Arul, Venkatesulu, 2012), where the pixel values of pair of adjacent pixels are shown as scatter plot (Figure 4).

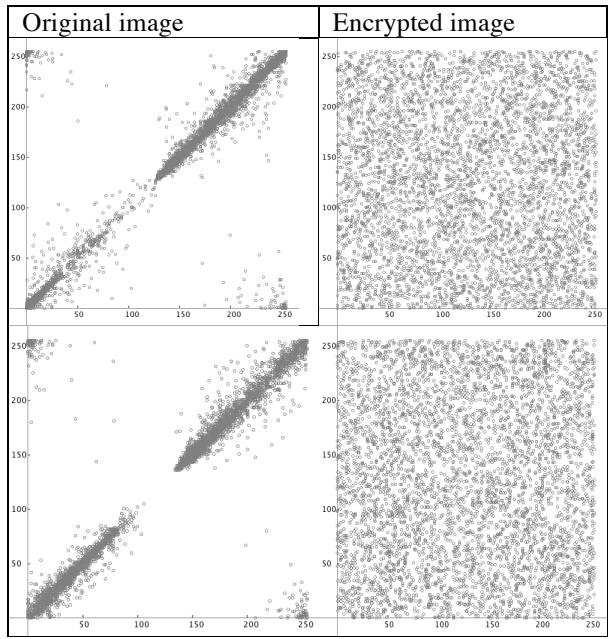

**Figure 4.** Adjacency based maps for images from figure 3.

We can see that the encrypted images demonstrate CC close to 0 as well as scatter plot is completely different after encryption and demonstrates a random pattern image.

## 3.3 Encryption time

Encryption time indicates the speed of encryption. We measured it by encrypting image files by various VSEM modules. Decryption time is approximately the same as encryption time as the decryption code uses practically the same operations. Examples of time measurements are presented in Table 1.

**Table 1.** The encryption time for processing image files

| Size | evsem_x | evsem_t | evsem_s | evsem_ct | all | AES |
|---|---|---|---|---|---|---|
| 280 Kb | 10 | 12 | 8 | 15 | 40 | 185 |
| 1Mb | 12 | 26 | 12 | 20 | 61 | 195 |
| 25Mb | 67 | 702 | 147 | 122 | 1049 | 405 |

It is appropriate to encrypt small records as well as typical photo images. For bigger files the speed of encryption will be comparable of faster than AES if we will use combination of any 3 VSEM modules except evsem_t or will encrypt big files as sequential blocks of a few Mb size applying all 4 encryption modules to each such block.

## 4 FENDOFF APPLICATION FOR COMPUTERS AND MOBILE DEVICES

Mobile phones are currently becoming the universal devices to access various services and data such as bank accounts, on-line stores, emails and etc. The users need to have dozens logins and passwords to access these services. It is crucial to have a good protection of this information and to be able easily access it.

### 4.1 FendOff for iOS devices

We have developed FendOff application for iOS devices. The main menu of the application is presented in Figure 5. The application can be downloaded from the iTunes store: (https://itunes.apple.com/us/app/fendoff/id970441536?mt=8&ign-mpt=uo%3D4). It uses VSEM encryption/decryption modules to secure text information such as passwords, bank or credit cards accounts data and etc. organized in separate categories as well as files from image archive (iOS photo album). The short description and architecture diagram of FendOff are presented in Figure 5.

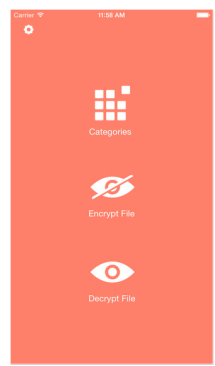

**Figure 5.** Main menu of the iOS FendOff application

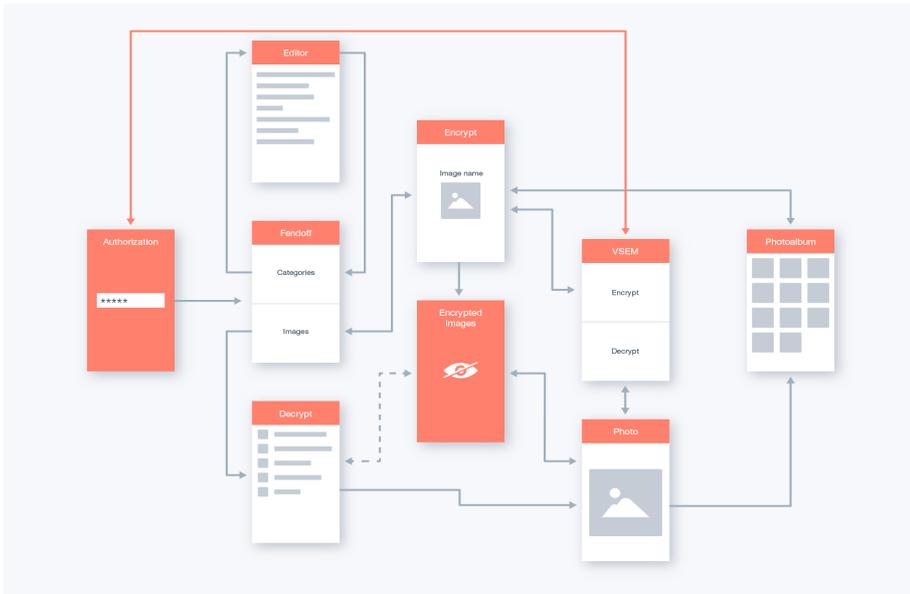

Figure 1 Diagram of FendOff iOS version. Successful authorization results in obtaining two structures: *Categories,* which maintains personal data stored in app and *Images,* which maintains encrypted images. Encryption occurs by choosing image from photo album, generating new password and storing encrypted image in special folder. All the information is then recorded in *Images* such as thumbnail, name and password. Decrypt allows to choose encrypted image, view it and if necessary save it to photo album.

## 4.2 FendOff for Android

Fendoff application was also implemented for Android. Unlike iOS version it supports encryption and decryption of all types of files due to access to the file system in this OS. The typical interaction with the application can be seen in Figure 7.

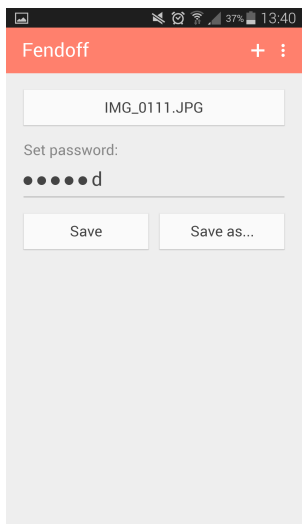

**Figure 7.** Encryption of file in FendOff Android application

## 4.3 FendOff for desktop computers

We created FendOff application versions for MAC and Windows OS to secure passwords type information and critical files. The Group/Categories panel of the application is presented in Figure 8 and informational panel for Computers login information can be seen in Figure 9.

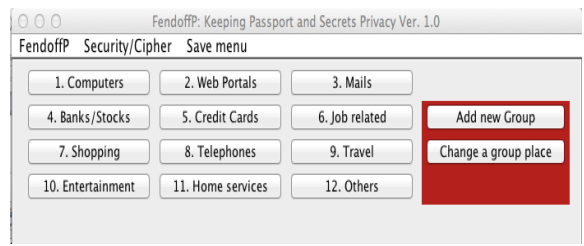

**Figure 8.** The Group/Categories panel of the FendOff application

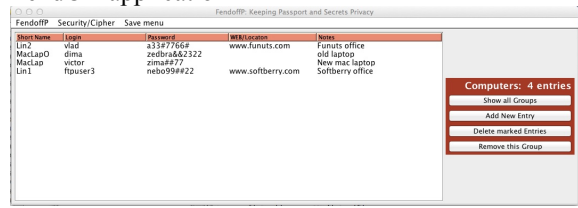

**Figure 9.** Informational panel for Computers login information of The FendOff application

## 5 CONCLUSIONS

This paper describes a design of a new encryption and decryption modules and their encryption quality and performance analysis. EQ and CC metrics indicate a good encryption quality of the implemented modules. The suggested encryption/and decryption algorithms have been applied in developing FendOff application to secure personal information on computers and mobile devices.